\documentclass[a4paper,11pt]{article}
\usepackage{pos}

\usepackage{xcolor}

\usepackage{booktabs}
\usepackage{multirow}
\usepackage{array}

\usepackage{float}
\graphicspath{{./figures/}}

\DeclareMathOperator{\diag}{diag}

\title{Spectral Diffusion for Sampling on ${\rm SU}(N)$}

\author[a]{Gurtej Kanwar}
\author*[b]{Octavio Vega}

\affiliation[a]{Higgs Centre for Theoretical Physics,
  University of Edinburgh, Edinburgh EH9 3FD, UK}

\affiliation[b]{Illinois Center for Advanced Studies of the Universe and Department of Physics,\\
University of Illinois Urbana-Champaign, Urbana, IL 61801, USA}

\emailAdd{octavio5@illinois.edu}

\abstract{Although ensemble generation remains a central challenge in lattice field theory simulations, recent advances in generative modeling may offer a path to accelerated sampling in these contexts. In this work, we implement a framework for efficiently training diffusion models acting on ${\rm SU}(N)$ degrees of freedom, adapting the traditional score matching technique to the group manifold. We demonstrate that our models can effectively reproduce several target densities, resulting in precise unbiased expectation values. These results mark a step for diffusion models towards modeling full ${\rm SU}(N)$ lattice field theories, including lattice Quantum Chromodynamics.}

\FullConference{The 42nd International Symposium on Lattice Field Theory (LATTICE2025)\\
2-8 November 2025\\
Tata Institute of Fundamental Research, Mumbai, India\\}

\begin{document}
\maketitle

\section{Introduction}
The first step of any lattice field theory simulation is ensemble generation, which requires accurate sampling of physical probability densities in high dimensions.
Traditional approaches, such as Hybrid Monte Carlo~\cite{Duane:1987de}, are afflicted by topological freezing~\cite{woit:1983, alles:1996} and critical slowing down~\cite{wolff:1990}, which become more pronounced with decreasing lattice spacing and increasing dimensionality.
Recently, several advancements in machine learning have offered methods for using deep generative models to efficiently produce samples from arbitrarily complicated probability densities. Among these, score-based diffusion models~\cite{sohl-dickstein:2015, ho:2020, song:2019, song:2021} have
shown promising impacts in image generation, drug discovery, and, recently, lattice field theory~\cite{wang:2023, zhu:2025, vega:2025}. This work extends the formulation of score-based diffusion models to ${\rm SU}(N)$ degrees of freedom, laying the groundwork for their application to lattice gauge theories, such as lattice Quantum Chromodynamics. While previous works have demonstrated a general construction of Riemannian manifold diffusion models~\cite{huang:2022, Lou:2023}, here we resolve several practical questions required to implement these models for ${\rm SU}(N)$. A software package containing this implementation is available at \href{https://github.com/ovega14/sun_diffusion}{\tt github.com/ovega14/sun\_diffusion}.

The central idea of score-based diffusion models is connecting the target probability distribution $p_0$ to a trivial distribution by incrementally applying Brownian noise. This \emph{diffusion process} induces a time-dependent sequence of distributions $p_t$ according to the associated Fokker-Planck equation; in practice a finite diffusion time $T$ is sufficient for $p_T$ to very closely approximate the equilibrium Gaussian (for non-compact variables) or uniform (for compact variables) distribution. The target distribution can be efficiently sampled using the reverse transport process, given by an ordinary differential equation written in terms of the \emph{score function} $\nabla \log p_t$. Because the score function is related to the unknown density $p_t$, score-based diffusion models instead optimize a machine learning approximation of the true score, $s_t \approx \nabla \log p_t$. Regardless, exact unbiased estimates can be obtained by constructing appropriate reweighting factors.\footnote{The learned reverse transport process is a continuous normalizing flow, and the unbiased estimators used in that context directly apply here~\cite{Gerdes:2022eve}.}

\section{Diffusion and the Heat Equation}
A fundamental step in training score-matching diffusion models is drawing samples over the range of diffusion times $t$ to be studied. Efficient training is possible when these samples can be drawn directly, without simulating the diffusion process.

\subsection{Euclidean Diffusion}
Mathematically, the diffusion and denoising processes are formulated through stochastic differential equations (SDE). A typical choice for the diffusion process on Euclidean data is the \emph{variance-expanding} SDE,\footnote{Here, $dW_t$ is an infinitesimal Wiener process increment defined by $dW_t = W_{t + dt} - W_t \sim \mathcal{N}(0, dt)$.}
\begin{equation}\label{eq:eucl_ve_sde}
    dx_t = g_t dW_t.
\end{equation}
The time-dependent probability density $p_t(x)$ of the samples evolves according to the Fokker-Planck Equation, which is in this case the heat equation
\begin{equation}\label{eq:eucl_heat_eqn}
    \partial_t p_t(x) = \frac{g_t^2}{2} \Delta p_t(x), \quad \Delta = \partial_x^2.
\end{equation}

The fundamental solution to Eq.~\eqref{eq:eucl_heat_eqn} is the \emph{heat kernel},
\begin{equation}
    K_t(x) = \tfrac{1}{\sqrt{2\pi \sigma_t^2}} \exp\left[ -\frac{1}{2\sigma_t^2}  \|x\|^2\right],
\end{equation}
where the diffusivity $\sigma_t \geq 0$ is defined by
$\sigma_t = \sqrt{\int_0^t g_{t'}^2 \, dt'}$.
The forward evolution of any target density $p_0(x)$
has an exact solution obtained by convolving against $K$:
\begin{align}
    p_0(x) &\mapsto p_t(x) = \int_{\mathbb{R}^d} K_t(x - y) p_0(y) \; dy .
\end{align}
Thus, simulating the forward diffusion process from $x_0$ to $x_t$ is equivalent to solving the heat equation to arbitrary time $t$. This requires sampling noise $\Delta x_t \sim K_t(\cdot)$, which is trivially accomplished in Euclidean space by sampling $\eta \sim \mathcal{N}(0, \mathbb{I})$ and setting $\Delta x_t = \sigma_t \eta$, leading to the forward map
\begin{align}
    x_0 &\mapsto x_t = x_0 + \sigma_t\eta.
\end{align}

\subsection{Diffusion Processes on SU(N)}
An analogous diffusion process can be defined on the ${\rm SU}(N)$ group manifold as
\begin{equation} \label{eq:sun_ve_sde}
    dU_t = i g_t \left[ \sum_{a=1}^{N^2 - 1} T^a dW_t^a \right] U_t.
\end{equation}
Due to the non-Abelian nature of ${\rm SU}(N)$, the stochastic evolution of $U_t$ is given by a path-ordered exponential $U_t = \mathcal{P}\exp\left[i \int_0^t g_t \sum_{a=1}^{N^2 - 1} T^a dW_t^a\right] U_0$. Sampling $U_t$ using simulation would thus require a computationally challenging numerical approximation to this integral. However, we demonstrate in the following section that the analytically known heat kernel of ${\rm SU}(N)$ can be sampled directly as an efficient alternative.

\section{The Heat Kernel over SU(N)}
The Fokker-Planck equation corresponding to~\eqref{eq:sun_ve_sde} is obtained by replacing the Euclidean Laplace operator with the Laplace-Beltrami operator in~\eqref{eq:eucl_heat_eqn}, giving
\begin{equation} \label{eq:sun_heat}
    \partial_t p_t(U) = \frac{g_t^2}{2} \Delta_U p_t(U),
    \quad \Delta_U f(U)= \sum_{a=1}^{N^2-1} \frac{\partial^2}{\partial \omega_a^2} f\left( e^{i \omega_a T^a} U \right)\Bigg|_{\omega=0}.
\end{equation}
The ${\rm SU}(N)$ heat kernel $K_t(U)$ is the Green's function solution of \eqref{eq:sun_heat} sourced by a delta distribution at the identity, so that forward evolution under the heat equation is solved by convolving against $K$,
\begin{equation}
    p_0(U) \mapsto p_t(U) = \int dV \; K_t(U V^\dagger) \, p_0(V).
\end{equation}

Combining left- and right-invariance of $\Delta_U$ with the invariance of the identity matrix $I$ under unitary conjugation, the heat kernel satisfies $K_t(U) = K_t(P^\dagger U P)$. The heat kernel is therefore completely specified by its value on the subgroup of diagonal matrices.\footnote{This subgroup is known as the maximal torus $\mathbb{T} \leq {\rm SU}(N)$.} Matrices belonging to this subgroup can be identified with a list of eigenangles,
\begin{equation}
    \Lambda = \diag(\lambda_1, \dots, \lambda_N) = \diag(e^{i \theta_1}, \dots, e^{i \theta_N}) \quad \leftrightarrow \quad \boldsymbol{\theta} = (\theta_1, \dots, \theta_N),
\end{equation}
where $\sum_j \theta_j \equiv 0 \pmod{2\pi}$. By convention, we chose $\theta_j \in [-\pi, \pi]$ for all explicit representations of the eigenangles below. With a slight abuse of notation, we write $K_t(\boldsymbol{\theta})$ for the heat kernel on the diagonal subgroup.

\subsection{Wrapped Normal Representation}
The heat kernel on the group is closely related to the heat kernel on the Cartan subalgebra of $\mathfrak{su}(N)$, which is isomorphic to the space $\{ \boldsymbol{x} \,|\, \sum_{i=1}^N x_i = 0 \}$. In particular,
applying an appropriate measure factor and wrapping using the exponential map, $\theta_j = \arg(e^{i x_j})$, leads to the representation~\cite{Menotti:1981ry}
\begin{equation}\label{eq:wrapped_normal_sun_hk}
\begin{aligned}
    K_t(\boldsymbol{\theta}) \propto \sum_{\boldsymbol{m} \in \mathbb{Z}^N} \tilde{K} _t(\boldsymbol{\theta} + 2\pi \boldsymbol{m}) \, \delta\left(\sum_i m_i\right), \quad 
    \tilde{K}_t(\boldsymbol{x}) = \prod_{i < j} \frac{x_i - x_j}{2 \sin\left[\frac{x_i - x_j}{2}\right]} \exp\left[-\frac{1}{2\sigma_t^2} \sum_i x_i^2\right],
\end{aligned}
\end{equation}
where the sum over $\boldsymbol{m} \in \mathbb{Z}^N$ represents the sum over all equivalent pre-images of $\boldsymbol{\theta}$.\footnote{The sums in Eq.~\eqref{eq:wrapped_normal_sun_hk} and Eq.~\eqref{eq:wrapped_score_fn} can be calculated more stably using the $\mathbf{LogSumExp}$ function to avoid singularities.}

We can also analytically compute the score function for~\eqref{eq:wrapped_normal_sun_hk}. First, the score for the heat kernel over the space of unwrapped eigenangles is given by
\begin{equation}\label{eq:unwrapped_sun_score_fn}
    \partial_i\log \tilde{K}_t(\boldsymbol{x}) = - \frac{x_i}{\sigma_t^2} + \sum_{j \neq i}\left[\frac{1}{x_i - x_j} - \frac{1}{2}\cot\frac{x_i - x_j}{2}\right] ,
\end{equation}
where the first term is the Euclidean score function for the Gaussian distribution and the second term corresponds to the gradient of the measure term.
The wrapped score is then
\begin{equation}\label{eq:wrapped_score_fn}
    \nabla \log K_t(\boldsymbol{\theta}) = \frac{1}{K_t(\boldsymbol{\theta})}\sum_{\boldsymbol{m} \in \mathbb{Z}^N} \tilde{K}_t(\boldsymbol{\theta} + 2\pi\boldsymbol{m})\nabla \log \tilde{K}_t(\boldsymbol{\theta} + 2\pi\boldsymbol{m}) \, \delta\left(\sum_i m_i\right).
\end{equation}

\subsection{Dual Representation}
The spectral heat kernel may also be written as a character sum or eigenfunction expansion which, for general $N$, takes the form~\cite{Menotti:1981ry, Baaquie:1988xk}
\begin{equation}\label{eq:sun_hk_character_expansion}
    K_t(U) = \sum_{\boldsymbol{\mu}}\dim(\mu) \chi_{\boldsymbol{\mu}}(U) \exp \left[ -\sigma_t^2 c_2(\boldsymbol{\mu}) \right]
\end{equation}
where the sum runs over partitions $\boldsymbol{\mu} = \{ \mu_1 \geq \dots \geq \mu_N = 0\}$ labeling the ${\rm SU}(N)$ irreps. The value of the quadratic Casimir is the eigenvalue of the ${\rm SU}(N)$ Laplace-Beltrami operator:
\begin{equation}
    c_2(\boldsymbol{\mu}) = \sum_{j=1}^N \mu_j (\mu_j - 2j + N + 1) - \frac{1}{N} \left(\sum_{j=1}^N \mu_j\right)^2.
\end{equation}
The Weyl character and Weyl dimension formulas~\cite{Drouffe:1983fv} give
\begin{equation}\label{eq:weyl_character_and_dimension_formula}
    \chi_{\boldsymbol{\mu}}(U) = \frac{\det A(U)}{\det B(U)}, \quad \dim(\boldsymbol{\mu}) = \prod_{j < k} \frac{\mu_j - \mu_k + k - j}{k - j}
\end{equation}
where the matrix elements of $A$ and $B$ are given in terms of the eigenvalues $\lambda_i$ of $U$ as
\begin{equation}
    [A(U)]_{ij} = \lambda_i^{\mu_j + N - j}, \quad [B(U)]_{ij} = \lambda_i^{N - j}.
\end{equation}
The resulting score function is then
\begin{equation}
    \partial_i \log K_t(\boldsymbol{\theta}) = \frac{1}{K_t(U)} \sum_{\boldsymbol{\mu}}\dim(\boldsymbol{\mu}) \chi_{\boldsymbol{\mu}}(U) {\rm Tr}\left[A^{-1}\partial_{\theta_i} A - B^{-1}\partial_{\theta_i} B\right] \exp\left[-\sigma_t^2 c_2(\boldsymbol{\mu})\right].
\end{equation}

In contrast to \eqref{eq:wrapped_normal_sun_hk}, here $\sigma_t^2$ appears in the numerator of the exponent. These two representations are complementary: if $\sigma_t$ is small, the sum over $\boldsymbol{m}$ rapidly converges in \eqref{eq:wrapped_normal_sun_hk}, while if $\sigma_t$ is large, the sum over $\boldsymbol{\mu}$ rapidly converges in \eqref{eq:sun_hk_character_expansion}.
For the limited range of $\sigma_t$ used in the one-variable targets below, we find that the wrapped normal representation remains stable for all times, but we expect that larger $\sigma_t$ may be required for higher-dimensional examples to fully diffuse the data to a uniform prior distribution, making the dual representation critical for stable training.

\subsection{Sampling the Heat Kernel}
Given the explicit representations above, the full heat kernel over the group can be efficiently sampled in a two-step process. First, we sample the \emph{marginal} distribution over eigenvalues,
\begin{equation}
    \boldsymbol{\theta} \sim \prod_{i < j} 4 \sin^2 \left[ \frac{\theta_i - \theta_j}{2} \right] K_t(\boldsymbol{\theta}),
\end{equation}
where marginalizing over the eigenvectors introduces the extra Haar measure term. Second, we sample eigenvectors $P \in {\rm U}(N)$ uniformly with respect to the Haar measure of ${\rm U}(N)$, and construct
\begin{equation} \label{eq:group-hk-sample}
    V = P \diag\left(e^{i\boldsymbol{\theta}}\right) P^\dagger \sim K_t(V).
\end{equation}
In this work, we sample the marginal distribution using a Metropolis sampling scheme with a proposal distribution tuned separately for $\sigma_t < 0.5$ (wrapped normal) and $\sigma_t > 0.5$ (uniform). A similar sampling approach was successfully applied in Ref.~\cite{Lou:2023}.

\section{Spectral Diffusion}
\begin{figure}[t]
    \centering
    \includegraphics[width=1.0\linewidth]{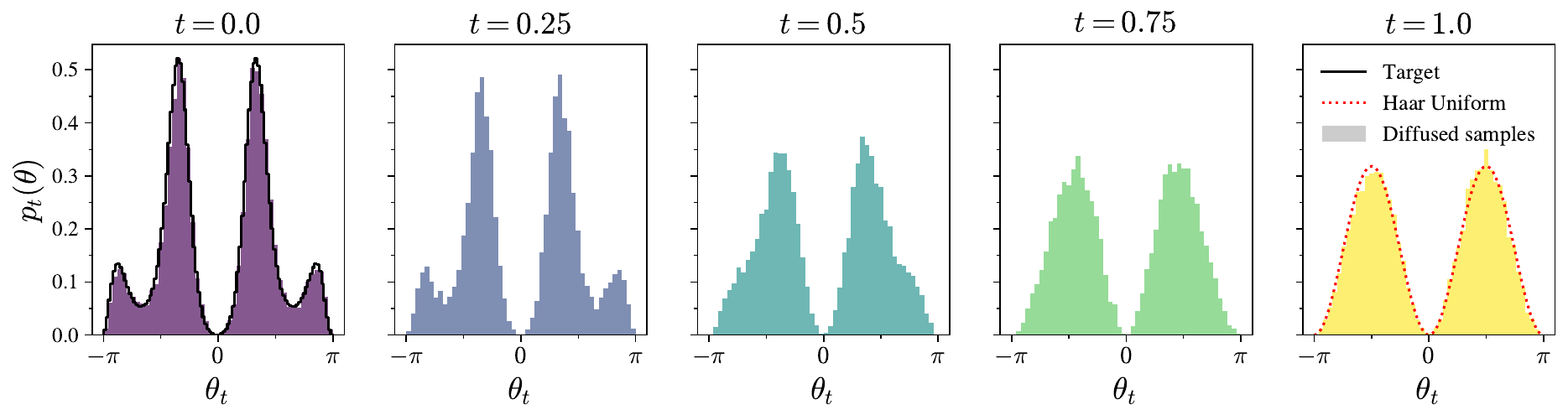}
    \caption{Evolution from left to right of the SU(2) spectral density on the first eigenangle under the forward diffusion process. The starting (target) density is progressively smoothed out by the heat kernel. After 1 unit of diffusion time, the diffused density closely matches the ${\rm SU}(2)$ Haar uniform density.}
    \label{fig:su2_fwd_diffusion}
\end{figure}
The forward diffusion process on the group is simulated by
\begin{equation}\label{eq:sun_fwd_diffusion}
U_0 \mapsto U_t = V \, U_0,
    \quad V = P \diag\left(e^{i \boldsymbol{\theta}}\right) P^\dagger \sim K_t(V).
\end{equation}
We show an example of the forward spectral diffusion process on ${\rm SU}(2)$ in Figure~\ref{fig:su2_fwd_diffusion}.

\subsection{Group-Valued Score Matching}
Score-based diffusion models are conventionally trained via \emph{score matching}, which minimizes the distance between the model score function and an unbiased estimator of the true score function across diffusion time. The ${\rm SU}(N)$ score-matching loss function, adapted to the ${\rm SU}(N)$ group manifold, is
\begin{equation}\label{eq:score_matching_loss_fn}
    \mathcal{L} = \int_0^1 \mathbb{E}_{U_0 \sim p_0} \mathbb{E}_{V \sim K_t} \left[\sigma_t^2\left\|s_t(U) - P^\dagger {\rm diag}[\nabla \log K_t(\boldsymbol{\theta})] P\right\|^2\right] dt.
\end{equation}
Here, $\boldsymbol{\theta}$ and $P$ are the sampled eigenangles and eigenvectors used to construct $V$ in \eqref{eq:group-hk-sample}, $s_t(V)$ takes values in the algebra $\mathfrak{su}(N)$, and the norm is over algebra components.
As usual for score matching, computing \eqref{eq:score_matching_loss_fn} requires samples $U_0$ from the target distribution.
To verify the correctness of our approach, we have applied the training routine detailed here to successfully learn the analytically known heat kernel score function \eqref{eq:wrapped_score_fn} by training with a fixed initial condition $U_0 = I$.

\subsection{Reverse Transport}
Given a trained score function, the ODE for the reverse transport process is
\begin{equation}
    \partial_t U_t = \left[ -\tfrac{i}{2} g_t^2 s_t(U_t) \right] U_t,
\end{equation}
to be integrated in reverse from $t = 1 \to 0$. We can compute the model probability density at each time $t$ as
\begin{equation}
    \log q_t (U_t) = \int_t^1 \frac{g_{t'}^2}{2} \nabla \cdot s_{t'}(U_{t'}) \, dt'.
\end{equation}
This allows reweighting factors to be introduced for unbiased evaluation at $t = 0$: for each sample $U$ drawn, we evaluate the model density $\log q(U) \equiv \log q_0(U)$ and the target density $\log p(U_i) = -S(U_i) + \text{const.}$, giving the unnormalized weights
\begin{equation}\label{eq:reweighting_factors}
    w(U) = e^{-S(U)} / q(U).
\end{equation}

\section{Toy Example}
To illustrate the ability of our method to sample from non-trivial target distributions, we define a family of simple densities $p^{(i)}$ over a single ${\rm SU}(N)$ matrix $U$, where the action for the toy theory is polynomial in the matrix:
\begin{equation}
    S^{(i)}(U) = -\frac{\beta}{N} {\rm Re}\,{\rm Tr}\left[\sum_{n=1}^3 c_n^{(i)} U^n  \right], \quad p^{(i)}(U) = \frac{1}{Z}e^{-S^{(i)}(U)}.
\end{equation}
The action is parameterized by a set of coefficients $c_n^{(i)}$, $i \in \{1, 2, 3\}$ which determine the shape of the density. For comparison to the spectral flow-based approach, we choose the same sets of coefficients as in Table I of Ref.~\cite{Boyda:2020hsi}. We investigate both ${\rm SU}(2)$ and ${\rm SU}(3)$ targets.

The diffusion process uses a diffusion schedule $g_t = \kappa t^\alpha$ with $\kappa=3.0$, $\alpha=1.0$, and we stop diffusion at $t=1$, which is sufficient to reach a uniform distribution.
For training, 32,768 configurations were generated using the Metropolis algorithm with 1000 thermalization steps, keeping every tenth configuration. Unless otherwise stated, we train for 100 epochs using the Adam optimizer~\cite{kingma:2015} with a batch size of 1024 and a base learning rate of $10^{-3}$; for SU(3) experiments we instead use a batch size of 3072 and a learning rate of $10^{-2}$. Using the fact that the target distribution is a function of only the eigenvalues, we define a specialized score network $s_t(U_t) = s_t(\boldsymbol{\theta}_t)$ acting directly on eigenvalues and returning non-zero values only in the Cartan subalgebra. The score network is parameterized by a multilayer perceptron with three hidden layers and \textbf{SiLU} activations, using sinusoidal positional encodings to embed the input times.

\begin{figure}[tp]
    \centering
    \includegraphics[width=1.0\linewidth]{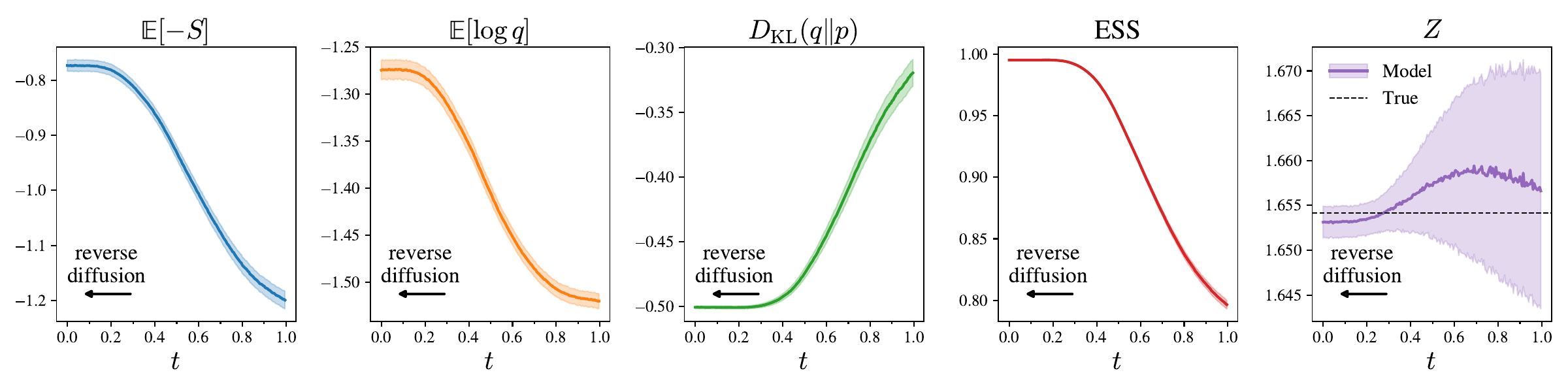}
    \caption{Reverse transport process for the toy ${\rm SU}(2)$ model with $\beta = 1.0$ and coefficient set $c^{(3)}$. The horizontal axis on each plot represents forward diffusion time, so the evolution should be interpreted as right to left. The KL divergence is seen to decrease while the ESS increases over time, signifying that the target density is being approximately reproduced by the model during sampling.}
    \label{fig:sampling_histories}
\centering
    \includegraphics[width=1.0\linewidth]{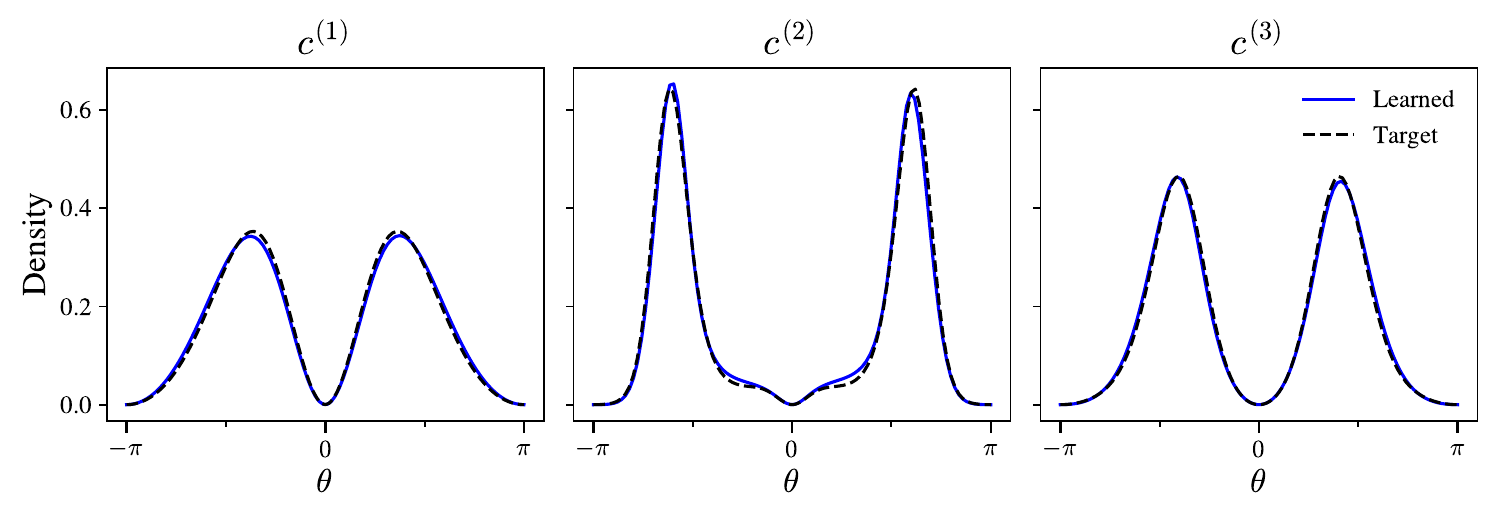}
    \caption{Learned and target ${\rm SU}(2)$ densities at $\beta = 1.0$. Models were trained for 200 epochs. Unbiased results can be obtained by reweighting estimates from the model to the target density with \eqref{eq:reweighting_factors}.}
    \label{fig:su2_final_densities}
\centering
    \includegraphics[trim={0 0.5cm 0 0.5cm},clip,width=0.8\linewidth]{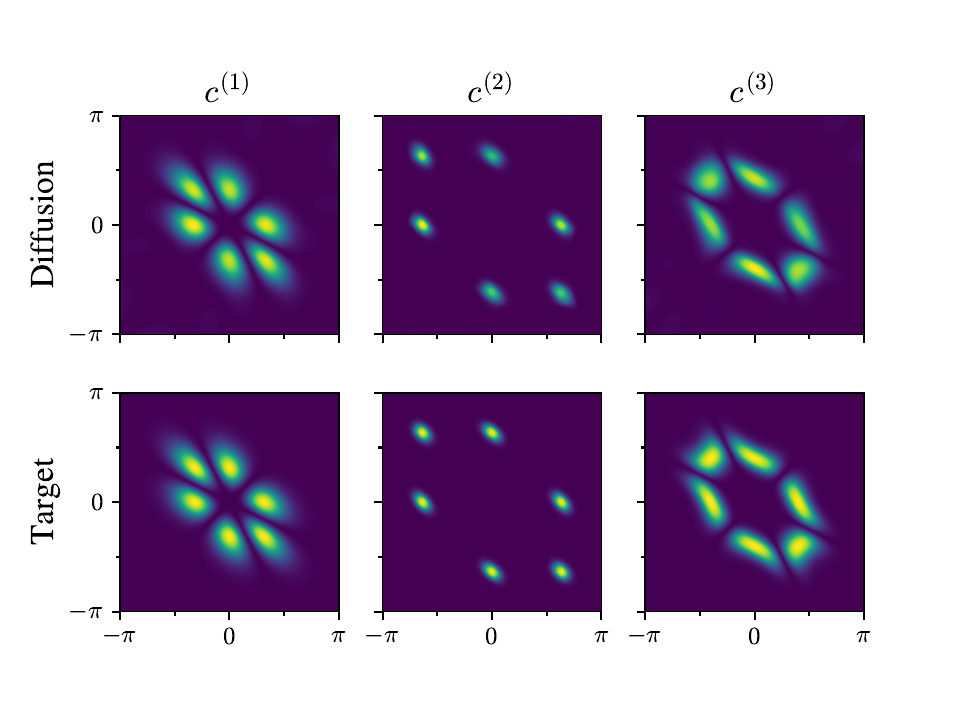}
    \caption{Comparison between the learned and target ${\rm SU}(3)$ densities at $\beta = 9.0$, shown over the first two eigenangles $\theta_1$ and $\theta_2$. Unbiased results can be obtained from the diffusion model with \eqref{eq:reweighting_factors}.} \label{fig:su3_final_densities}
\end{figure}

For inference, we sample the uniform distribution and apply reverse transport using a simple group-valued Euler integrator with 200 steps. Fig.~\ref{fig:sampling_histories} displays the measure target and model log-likelihoods, $-S$ and $\log q$, the unnormalized reverse Kullback-Leibler divergence $D_{\rm KL}\left(q \| p\right) \equiv \mathbb{E}_{U \sim q}[\log{q} + S]$
, the effective sample size (ESS),
\begin{equation}
    {\rm ESS} := \frac{\mathbb{E}\left[w(U)\right]^2}{\mathbb{E}\left[w(U)^2\right]}, \quad 0 \leq {\rm ESS} \leq 1, \end{equation}
as well as an unbiased estimator of the partition function, $Z = \int dU \, e^{-S(U)} = \mathbb{E}[w]$.
We summarize the final effective sample sizes we obtain for the toy models in Table~\ref{tab:su2_toy_ess} and compare the learned distributions with the target densities in Figs.~\ref{fig:su2_final_densities} and \ref{fig:su3_final_densities}.
\begin{table}[t]
\centering
\setlength{\tabcolsep}{4pt} \renewcommand{\arraystretch}{1.2}
\begin{tabular}{@{} l @{\hspace{20pt}} c c c @{\hspace{20pt}} c c c @{\hspace{20pt}} c c c @{}}
    \toprule
    & \multicolumn{1}{c}{} & \multicolumn{1}{c}{$c^{(1)}$} & \multicolumn{1}{c}{} 
    & \multicolumn{1}{c}{} & \multicolumn{1}{c}{$c^{(2)}$} & \multicolumn{1}{c}{} 
    & \multicolumn{1}{c}{} & \multicolumn{1}{c}{$c^{(3)}$} & \multicolumn{1}{c}{} \\
    \midrule
    $\beta$  & 1.0 & 5.0 & 9.0  
              & 1.0 & 5.0 & 9.0  
              & 1.0 & 5.0 & 9.0 \\
    ${\rm SU}(2)$ ESS (\%)  & 99.6 & 96.6 & 97.1 
                            & 96.7 & 97.8 & 95.4 
                            & 99.5 & 98.4 & 98.7 \\
    ${\rm SU}(3)$ ESS (\%)  & 98.8 & 93.7 & 97.7 
                            & 97.6 & 95.1 & 93.8 
                            & 99.2 & 97.1 & 96.8 \\
    \bottomrule
\end{tabular}
\caption{Final ESS values for different $\beta$ values and coefficient sets in the ${\rm SU}(N)$ toy model.}
\label{tab:su2_toy_ess}
\end{table}

\section{Outlook}
We have demonstrated an implementation of the ${\rm SU}(N)$ heat kernel, its score function, and training using ${\rm SU}(N)$ score matching. Analytic results for the score function of the heat kernel and a numerically stable implementation were key steps to making this method possible. This lays the groundwork for future studies  with the potential to address the challenges of ensemble generation for ${\rm SU}(N)$ lattice gauge theories.

\acknowledgments
We thank Diaa Habibi for bringing our attention to related results~\cite{Lou:2023} during the preparation of this work. We are grateful to the MITP Summer School 2025 for hosting an engaging exchange leading to the ideas presented here.
OV acknowledges support from a UIUC Graduate College Fellowship and from a Sloan Scholarship through the Alfred P. Sloan Foundation.
Numerical experiments and data analysis were executed using \textbf{PyTorch}~\cite{paszke:2019} and \textbf{NumPy}~\cite{harris:2020}. Figures were created with \textbf{Matplotlib}~\cite{hunter:2007}.
Some of the experiments in this research used the DeltaAI advanced computing and data resource, which is supported by the National Science Foundation (award OAC 2320345) and the State of Illinois.


\begin{thebibliography}{99}



\bibitem{Duane:1987de}
S.~Duane, A.~D.~Kennedy, B.~J.~Pendleton, and D.~Roweth,
\emph{Hybrid Monte Carlo},
\href{https://doi.org/10.1016/0370-2693(87)91197-X}{\emph{Phys. Lett. B} \textbf{195} (1987) 216}.

\bibitem{woit:1983}
P.~Woit,
\emph{Topological Charge in Lattice Gauge Theory},
\href{DOI: https://doi.org/10.1103/PhysRevLett.51.638}{\emph{Phys. Rev. Lett.} \textbf{51} (1983) 638}.

\bibitem{alles:1996}
B.~Allés, G.~Boyd, M.~D'Elia, A.~Di Giacomo, and E.~Vicari,
\emph{Hybrid Monte Carlo and topological modes of full QCD},
\href{https://doi.org/10.1016/S0370-2693(96)01247-6}{\emph{Phys. Lett. B} \textbf{389} (1996) 107}
[\href{https://arxiv.org/abs/hep-lat/9607049}{\tt hep-lat/9607049}].

\bibitem{wolff:1990}
U.~Wolff,
\emph{Critical Slowing Down},
\href{https://doi.org/10.1016/0920-5632(90)90224-I}{\emph{Nucl. Phys. B Proc. Suppl.} \textbf{17} (1990) 93}.

\bibitem{sohl-dickstein:2015}
J.~Sohl-Dickstein, E.~Weiss, N.~Maheswaranathan, S.~Ganguli,
\emph{Deep Unsupervised Learning using Nonequilibrium Thermodynamics},
\href{https://proceedings.mlr.press/v37/sohl-dickstein15.html}{\emph{PMLR} \textbf{37} (2015) 2256}
[\href{https://arxiv.org/abs/1503.03585}{\tt 1503.03585}].

\bibitem{ho:2020}
J.~Ho, A.~Jain, and P.~Abbeel,
\emph{Denoising diffusion probabilistic models}, 
\href{https://dl.acm.org/doi/abs/10.5555/3495724.3496298}{\emph{NIPS '20: Proceedings of the 34\textsuperscript{th} International Conference on Neural Information Processing Systems} \textbf{574} (2020) 6840}
[\href{https://arxiv.org/abs/2006.11239}{\tt 2006.11239}].

\bibitem{song:2019}
Y.~Song and S.~Ermon, 
\emph{Generative modeling by estimating gradients of the data distribution},
\href{https://dl.acm.org/doi/10.5555/3454287.3455354}{\emph{Proceedings of the 33\textsuperscript{rd} International Conference on Neural Information Processing Systems} \textbf{1067} (2019) 11918}
[\href{https://arxiv.org/abs/1907.05600}{\tt 1907.05600}].

\bibitem{song:2021}
Y.~Song, J.~Sohl-Dickstein, D.~P.~Kingma, A.~Kumar, S.~Ermon and B.~Poole, 
\emph{Score-Based Generative Modeling through Stochastic Differential Equations},
\href{https://arxiv.org/abs/2011.13456}{\tt 2011.13456}.

\bibitem{wang:2023}
L.~Wang, G.~Aarts and K.~Zhou, 
\emph{Diffusion models as stochastic quantization in lattice field
theory}, 
\href{https://link.springer.com/article/10.1007/JHEP05(2024)060}{\emph{JHEP} \textbf{05} (2024) 060}
[\href{https://arxiv.org/abs/2309.17082}{\tt 2309.17082}].

\bibitem{zhu:2025}
Q.~Zhu, G.~Aarts, W.~Wang, K.~Zhou, and L.~Wang,
\emph{Physics-Conditioned Diffusion Models for Lattice Gauge Theory},
\href{https://arxiv.org/abs/2502.05504}{\tt 2502.05504}.

\bibitem{vega:2025}
O.~Vega, J.~Komijani, A.~X.~El-Khadra, and M.~Marinkovic,
\emph{Group-Equivariant Diffusion Models for Lattice Field Theory},
\href{https://arxiv.org/abs/2510.26081}{\tt 2510.26081}.

\bibitem{huang:2022}
C.~Huang, M.~Aghajohari, A.~J.~Bose, P.~Panangaden, and A.~Courville, 
\emph{Riemannian Diffusion Models}, 
\href{https://dl.acm.org/doi/10.5555/3600270.3600469}{\emph{NIPS'22: Proceedings of the 36\textsuperscript{th} International Conference on Neural Information Processing Systems} \textbf{199} (2022) 2750}
[\href{https://arxiv.org/abs/2208.07949}{\tt 2208.07949}].

\bibitem{Lou:2023}
A.~Lou, M.~Xu, A.~Farris, and S.~Ermon,
\emph{Scaling Riemannian Diffusion Models},
\href{https://dl.acm.org/doi/10.5555/3666122.3669641}{\emph{NIPS '23: Proceedings of the 37\textsuperscript{th} International Conference on Neural Information Processing Systems} \textbf{3519} (2023) 80291}
[\href{https://arxiv.org/abs/2310.20030}{\tt 2310.20030}].

\bibitem{Gerdes:2022eve}
M.~Gerdes, P.~de Haan, C.~Rainone, R.~Bondesan and M.~C.~N.~Cheng,
\emph{Learning lattice quantum field theories with equivariant continuous flows},
\href{https://scipost.org/10.21468/SciPostPhys.15.6.238}{\emph{SciPost Phys.} \textbf{15} (2023) 238}
[\href{https://arXiv.org/abs/2207.00283}{\tt 2207.00283}].









\bibitem{Menotti:1981ry}
P.~Menotti and E.~Onofri,
\emph{The Action of $SU(N)$ Lattice Gauge Theory in Terms of the Heat Kernel on the Group Manifold},
\href{https://doi.org/10.1016/0550-3213(81)90560-5}{\emph{Nucl. Phys. B} \textbf{190} (1981) 288}.

\bibitem{Baaquie:1988xk}
B.~E.~Ehsan,
\emph{Character functions of ${\rm SU}(3)$},
\href{https://doi.org/10.1088/0305-4470/21/11/022}{\emph{J. Phys. A} \textbf{21} (1988) 2651}.





\bibitem{Drouffe:1983fv}
J.~M.~Drouffe and J.~B.~Zuber,
\emph{Strong Coupling and Mean Field Methods in Lattice Gauge Theories},
\href{https://doi.org/10.1016/0370-1573(83)90034-0}{\emph{Phys. Rept.} \textbf{102} (1983) 1}.

\bibitem{Boyda:2020hsi}
D.~Boyda, G.~Kanwar, S.~Racani{\`e}re, D.~J.~Rezende, M.~S.~Albergo, K.~Cranmer, D.~C.~Hackett and P.~E.~Shanahan,
\emph{Sampling using ${\rm SU}(N)$ gauge equivariant flows},
\href{https://doi.org/10.1103/PhysRevD.103.074504}{\emph{Phys. Rev. D} \textbf{103} (2021) 074504}
[\href{https://arxiv.org/abs/2008.05456}{\tt 2008.05456}].

\bibitem{kingma:2015}
D.~P.~Kingma and J.~Ba,
\emph{Adam: A Method for Stochastic Optimization},
\href{https://arxiv.org/abs/1412.6980}{\tt 1412.6980}.

\bibitem{paszke:2019}
A.~Paszke, S.~Gross, F.~Massa, A.~Lerer, J.~Bradbury, G.~Chanan et~al., 
\emph{PyTorch: An Imperative Style, High-Performance Deep Learning Library}, 
\href{https://dl.acm.org/doi/10.5555/3454287.3455008}{\emph{Proceedings of the 33\textsuperscript{rd} International Conference on Neural Information Processing Systems} \textbf{721} (2019) 8024}
[\href{https://arxiv.org/abs/1912.01703}{\tt 1912.01703}].

\bibitem{harris:2020}
C.~R.~Harris, K.~J.~Millman, S.~J.~van der Walt, R.~Gommers, P.~Virtanen, D.~Cournapeau
et~al., 
\emph{Array programming with NumPy}, 
\href{https://doi.org/10.1038/s41586-020-2649-2}{\emph{Nature} \textbf{585} (2020) 357}
[\href{https://arxiv.org/abs/2006.10256}{\tt 2006.10256}].



\bibitem{hunter:2007}
J.~D.~Hunter,
\emph{Matplotlib: A 2D Graphics Environment},
\href{https://doi.org/10.1109/MCSE.2007.55}{\emph{Computing in Science \& Engineering} \textbf{9} (2007) 90}.



\end{thebibliography}
\end{document}